\documentclass[aps,preprintnumbers,superscriptaddress,groupedaddress]{revtex4}
\usepackage{graphicx}
\usepackage{amsmath}
\usepackage{dcolumn}   
\usepackage{bm}        
\usepackage{amssymb}   
\usepackage{color}
\usepackage{multirow}
\usepackage{mathtools}

\newcommand{\ket}[1] {\vert{#1}\rangle}

\newcommand{\be}{\begin{equation}}
\newcommand{\ee}{\end{equation}}
\def\ba{\begin{eqnarray}}
\def\ea{\end{eqnarray}}

\usepackage{xcolor}
\usepackage{multirow,booktabs}
\usepackage{tabularx}

\newcolumntype{C}[1]{>{\centering\let\newline\\\arraybackslash\hspace{0pt}}m{#1}}

\AtBeginDocument{
\heavyrulewidth=.08em
\lightrulewidth=.05em
\cmidrulewidth=.03em
\belowrulesep=.65ex
\belowbottomsep=0pt
\aboverulesep=.4ex
\abovetopsep=0pt
\cmidrulesep=\doublerulesep
\cmidrulekern=.5em
\defaultaddspace=.5em
}

\begin{document}

\title{Maximal Entanglement in High Energy Physics}

\author{Alba Cervera-Lierta}
\affiliation{Dept. F\'isica Qu\`antica i Astrof\'isica, Universitat de Barcelona, Diagonal 645, 08028 Barcelona, Spain.\\} 
\author{Jos\'e I. Latorre}
\affiliation{Dept. F\'isica Qu\`antica i Astrof\'isica, Universitat de Barcelona, Diagonal 645, 08028 Barcelona, Spain.\\} 
\affiliation{Center for Quantum Technologies, National University of Singapore.} 
\author{Juan Rojo}
\affiliation{Department of Physics and Astronomy, VU University, De Boelelaan 1081, 1081HV Amsterdam, \\
and Nikhef, Science Park 105, NL-1098 XG Amsterdam, The Netherlands\\}
\author{Luca Rottoli}
\affiliation{Rudolf Peierls Centre for Theoretical Physics, University of Oxford, 1 Keble Road, Oxford OX1 3NP, UK.\\}
\date{\today}
\maketitle


\section*{Abstract}
We analyze how maximal entanglement is generated at the fundamental level in QED by studying correlations between helicity states in tree-level scattering processes at high energy. We demonstrate that two mechanisms for the generation of maximal entanglement are at work: {\sl i)} $s$-channel processes where the virtual photon carries equal overlaps of the helicities of the final state particles, and {\sl ii)} the indistinguishable superposition between $t$- and $u$-channels. We then study whether {\it requiring} maximal entanglement constrains the coupling structure of QED and the weak interactions. In the case of photon-electron interactions unconstrained by gauge symmetry, we show how this requirement allows to reproduce QED. For $Z$-mediated weak scattering, the maximal entanglement principle leads to non-trivial predictions for the value of the weak mixing angle $\theta_W$. Our results are a first step towards understanding the connections between maximal entanglement and the fundamental symmetries of high-energy physics.


\section{Introduction}
\label{sec:intro}

Entanglement \cite{schrodinger} is the key property that pervades many developments in quantum physics.
As a paramount example, entanglement is necessary to discriminate classical from quantum physics using Bell inequalities~\cite{bell}. Entanglement can also be understood as the resource that enables genuine quantum protocols such as cryptography based on Bell inequalities~\cite{ekert} and teleportation~\cite{teleportation}. Large entanglement is expected to be present in quantum registers when a quantum algorithm produces a relevant advantage in performance over a classical computer such as Shor's algorithm~\cite{shor}. Entanglement also plays a crucial role in condensed matter, where quantum phase transitions in spin chains are characterized by a enhanced logarithmic scaling of entanglement entropy~\cite{conformal},
highlighting the relation between entanglement and conformal symmetry.

It is clear that entanglement is at the core of understanding and exploiting quantum physics. It is therefore natural to analyze the generation of entanglement at its most
fundamental origin,
namely the theories of fundamental interactions in particle physics.
If the quantum theory of electromagnetism, QED, would never generate entanglement among electrons,
Nature would never display a violation of a Bell inequality.
This implies that
entanglement must be generated by quantum unitary evolution at the fundamental level.

A deeper question emerges in the context of high-energy
physics. Is maximal entanglement (MaxEnt) possible at all?
In other words, are the laws of Nature such that MaxEnt
can always be realized?
One can imagine a QED-like theory where entanglement could be generated,
but in a way which would be insufficient to violate Bell inequalities.
Then, it would be formally possible to think of the existence of an underlying theory of hidden variables.
On the other hand, if MaxEnt is realized in QED, it is then possible to design
experiments that will discard classical physics right at the level of the scattering of elementary particles.
Taking a step further, one can ask what are the consequences of imposing that the laws of Nature are able to realize maximally entangled states. Can this requirement be promoted to a principle, and
to which extent is this principle consistent with fundamental symmetries such as gauge invariance?

The quest for simple postulates to describe  the fundamental
interactions observed in Nature resulted in the common acceptance of the
gauge principle, that is, the invariance of the physical laws over
internal local rotations for specific symmetry groups.
Leaving aside quantum gravity, 
the Standard Model describes electroweak and strong interactions 
by means of a Lagrangian which is largely dictated by gauge symmetry requirements.
It is natural to pursue further the search for yet an even simpler principle.
A possible candidate to formulate a basic underlying
axiom for local symmetries is provided by Information Theory.
We may recall the words of J. A. Wheeler, {\it ``all things physical are information-theoretic in origin"}
that substantiate his philosophy of {\sl ``it from bit''}~\cite{wheeler,wheeler2}.
Indeed, it is
conceivable that our equations are just a set of operations to 
implement basic information ideas and protocols.
Moreover, quantum physics is the natural candidate to build the ultimate computing device~\cite{feynman}.

The exploration of a concrete example of the {\sl ``it from bit''}
philosophy based on a maximal entanglement principle is the content of this work.

We shall show first than in QED only two mechanisms can generate MaxEnt in
high-energy scattering of fermions prepared in an initial helicity product state.
These are {\it i)} $s$-channel processes where the
virtual photon carries equal overlaps of the helicities of the final state particles,
and {\it ii)}
processes which display interference between $t$ and $u$ channels.
  We will then illustrate the deep connection between maximal entanglement and the structure
  of the electron-photon interaction vertex in QED. Indeed, maximal entanglement in most channels is related to the exact form of the QED vertex. As a consequence, imposing that the laws of nature are able to deliver maximal entanglement is tantamount to imposing the QED vertex.
 We shall finally analyze the consequences of imposing MaxEnt on the weak interactions and
 discover some surprising constraints on the parameters of the Standard Model.

Let us notice that in two-particle systems the concepts of maximal entanglement and maximal entropy are equivalent, as the reduced density matrix for any of the two particles is proportional to the identity, which is the maximally entropic state.
For systems with more than two particles, the classification of entanglement becomes richer and does not necessarily correspond to the entropy of the subsystem. 
As in this work we focus only on processes with two particles, we shall use MaxEnt to refer indistinctly to maximal entanglement or maximal entropy.

 Some previous works have studied the role of entanglement in particle physics.
In Ref.~\cite{alp} it was shown that orthopositronium can decay into 
3-photon states that can be used to perform Bell-like experiments that discard classical physics faster than the standard 2-particle Bell inequality.
Bell inequalities have also been 
discussed in kaon physics~\cite{Kaons} and its relation with the
characterization of $T$-symmetry violation~\cite{bernabeu}
as well as in neutrino oscillations~\cite{neutrinos}.
How entanglement varies in an elastic scattering process has been
studied using the $S$-matrix formalism in~\cite{EntScatt}.
Note also recent work on entanglement in Deep Inelastic Scattering~\cite{DIS}.
Also, a discussion
of quantum correlations in the CMB radiation has been brought to the domain of
Bell inequalities~\cite{maldacena}.

The outline of this paper is as follows.
In Sect.~\ref{sec:entanglement} we introduce measures to quantify entanglement
in scattering processes.
Then in Sect.~\ref{sec:maxentQED} we study how entanglement is generated in QED scattering
processes.
In Sect.~\ref{sec:maxent} we investigate to which extent
MaxEnt can be used as constraining principle on the structure of the
QED interactions.
Finally, in Sect.~\ref{sec:maxentweak} we assess some of the implications
of MaxEnt for the weak interactions, and in Sect.~\ref{sec:summary} we conclude.
A number of technical details of the calculations are collected in the appendices:
App.~\ref{app:QED}, about QED scattering amplitudes with helicity
dependence; App.~\ref{app:unconstrainedQED}, about unconstrained QED;
and App.~\ref{app:Z}, about helicity-dependent calculations in electroweak theory.

\section{Quantifying entanglement}
\label{sec:entanglement}

We shall study here scattering processes involving fermions and photons as incoming and outgoing particles. In the case of fermions (photons) we shall analyze the  entanglement of their helicities
(polarizations).
In both cases, the associated Hilbert space
is two-dimensional, and we use $\ket{0}$ and $\ket{1}$ as basis vectors.
The quantum state of an incoming or outgoing particle can then be written as
\be
  \label{genericstate}
  \ket{\phi}=\alpha \ket{00} + \beta \ket{01} +\gamma \ket{10} + \delta \ket{11}\, ,
\ee
where $|\alpha|^ 2+ |\beta|^ 2+ |\gamma|^ 2+ |\delta|^ 2=1$ due to normalization.
We note that in high-energy scattering a generic outgoing state 
will involve all possible outcomes of the process being analyzed.
The reduction to a two-level system therefore corresponds to a post-selection of results.
This is the correct description that delivers the probabilities which we could insert into a Bell inequality, once the final state has been identified.

To quantify entanglement we could use the Von Neumann entropy, but
this is not necessary since in our case 
all possible entanglement measures are related to 
\be
  \label{concurrence}
  \Delta \equiv 2 \left| \alpha\delta - \beta \gamma \right| \, ,
\ee
known as the {\sl concurrence}~\cite{concurrence}.
By construction,
$0 \le \Delta \le 1$, where the extreme cases $\Delta=0~(1)$ correspond to a product (maximally
entangled) state.

Here we shall study scattering process where the incoming particles are in a product state of their helicities, that is, the incoming particles are not entangled ($\Delta=0$).
It is certainly true that entanglement may emerge between other quantum numbers, such
as helicity and momentum. We here consider the case of helicities at any outgoing momenta.
This allows for analyzing entanglement in a wide range of physical distinct scatterings. 
The results are rich and already deliver information. A more sophisticated analysis should
include all possible quantum numbers, including momenta, flavor and color.
We will often work in the high-energy limit where helicities and chirality are equivalent,
and we will use the basis $\ket{0}_{\rm in}=\ket{R}$ and $\ket{1}_{\rm in}=\ket{L}$,
where $R$ and $L$ correspond  to right- and left-handed helicities, respectively. 
In general, the outgoing state will be a superposition of all possible helicity
combinations, and thus
the scattering amplitude of {\sl e.g.} $RL$ initial helicities, ${\cal M}_{RL}$, will
include each possible combination of outgoing helicities.
We will then parametrize scattering amplitudes as
\be
   {\cal M}_{RL} \sim \alpha_{RL}  \ket{RR}+\beta_{RL}  \ket{RL}+\gamma_{RL}
   \ket{LR}+\delta_{RL} \ket{LL} \, ,
\ee
where here the subscript in the right-hand side indicates the incoming helicities,
and the kets in the left-hand side indicate the values of the outgoing ones.\\
\vspace{-0.3cm}

\section{MaxEnt generation in QED}
\label{sec:maxentQED}

\begin{figure}[t]
\centering
\includegraphics[width=0.4\textwidth]{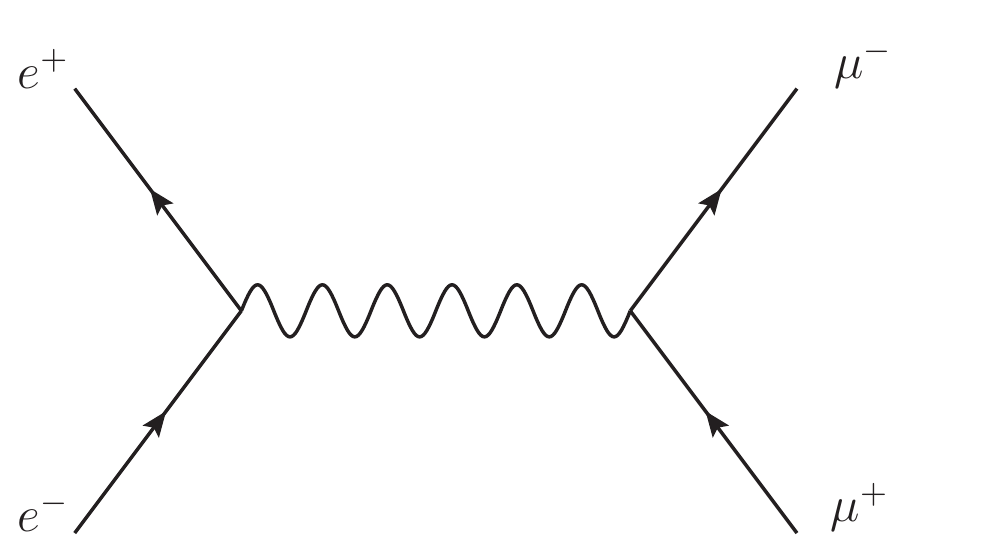}\\
\vspace{-0.2cm}
\caption{\small Feynman diagram for electron-positron scattering into
a muon-antimuon pair at tree level.
 }
  \label{Fig:epem_to_mupmum}
\end{figure}

Let us start our discussion with the analysis of how entanglement is generated in  electron-positron annihilation into a muon-antimuon pair, $e^ - e^ + \to \mu^ - \mu^ +$, described at tree level in QED by a single $s$-channel diagram (Fig. \ref{Fig:epem_to_mupmum}). 
In order to analyze entanglement, it is necessary to retain all the helicities
in the calculation.
As in the rest of this paper, we will work on the center-of-mass frame.

It is convenient to first focus on the current generated at the interaction vertices.
If the incoming particles propagate along the $z$-direction, 
the incoming current associated to two incoming particles in a $RL$ helicity product state
will be $\bar v_\uparrow \gamma^ \mu u_\uparrow = 2 p_0 (0, 1,i,0)$,
where $p_0$ is the electron's energy.
The outgoing particles will then be described by a current
as a function of $\theta$, the scattering angle.
As shown in appendix~\ref{app:QED},
we find that at high energies
the leading contribution only appears for incoming $RL$ (and $LR$) helicities,
\be
  \label{e+e-tomu+mu-}
  {\cal M}_{RL}\sim (1+\cos\theta) \ket{RL} +(-1+\cos\theta)\ket{LR} 
\ee
up to a prefactor which is not relevant here.
Therefore, for a scattering angle $\theta=\pi/2$
the final state becomes maximally entangled and proportional to $\ket{RL}-\ket{LR}$, with
$\Delta_{RL}=1$.
This result illustrates how MaxEnt can be generated
in a high-energy scattering process.
While scattering amplitudes in general carry a non-trivial angular dependence,
it is always possible to perform the measurement in the specific direction where MaxEnt is obtained,
not unlike the way maximally entangled states are obtained in quantum optics by parametric down conversion.
Let us also note that the dominant terms in the $e^ - e^ + \to \mu^- \mu^+$ scattering  at high energies are easily described by chirality conservation. This is not the case at lower energies,
where the emergence of entanglement is more complex. 

For incoming particles in the $RR$ helicity product state, all terms in the amplitude are suppressed by a power of $p_0$ as compared to the $RL$ case.
Nevertheless, MaxEnt is found for every angle $\theta$ and incoming momenta $p^0$.
An experiment that prepares $RR$ incoming states will therefore always result in MaxEnt.

It is instructive to revisit the computation of the $RL$ case focusing on the currents associated
to the virtual photon.
The incoming current (in the $z$-direction) corresponds to $j^{\mu\,(RL)}_{\rm in}= 2 p_0 (0,1,i,0)$,
and at high energies the non-vanishing outgoing currents at $\theta=\pi/2$ read 
$j^{\mu\,(RL)}_{\rm out}= 2 p_0 (0,0,-i,-1)$ and $j^{\mu\,(LR)}_{\rm out}= 2 p_0 (0,0,i,-1)$. Thus the third component of
$j_{\rm in}$ carries equal overlap (with different sign) of the two possible helicity
combinations for the outgoing state. In a sense, the photon cannot distinguish between those two options.
This is the basic element that leads to MaxEnt generation in $s$-channel processes.

\begin{figure}[t]
\centering
  {	\includegraphics[width=.25\columnwidth]{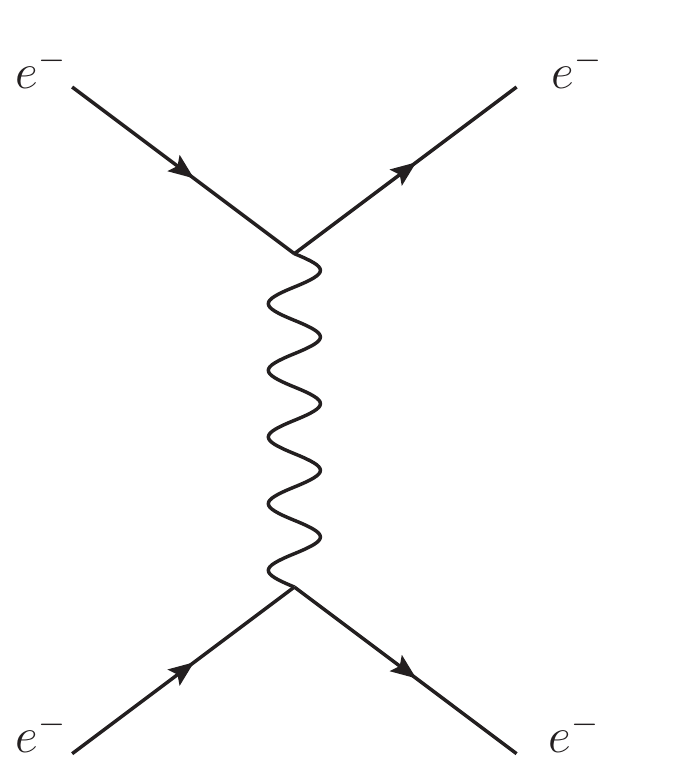}}
  {\includegraphics[width=.25\columnwidth]{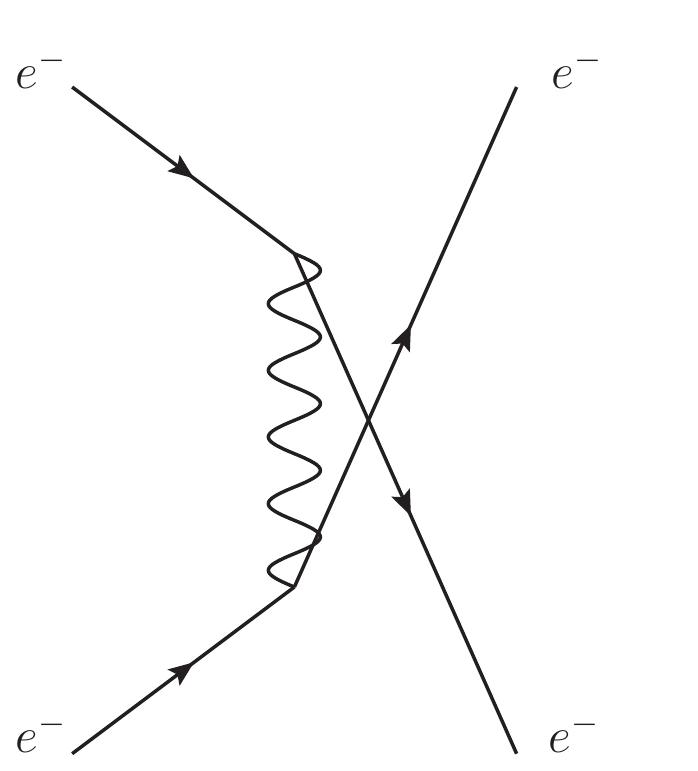}}\\
\vspace{-0.2cm}
\caption{\small  Feynman diagrams for
  M{\o}ller scattering, $e^-e^-\to e^-e^-$,
  in the   $t$ (left) and $u$ (right) channels.
  }
  \label{Fig:moeller}
\end{figure}

Entanglement can also be generated in QED through a completely different mechanism. Let us consider M\o ller (electron-electron elastic) scattering,
which receives contributions only from $t$- and $u$-channel diagrams (Fig.~\ref{Fig:moeller}).
For this process, the computation of the amplitude shows that no entanglement is generated
at high energies within each $t$ or $u$ channel separately, and that
the only entangled state is produced by their superposition, resulting in
${\cal M}_{RL} \sim (t/u) \ket{LR} - (u/t) \ket{RL}$ ,
leading to a concurrence
\be   
  \label{concurrenceee}
   \Delta_{RL} \xrightarrow[]{{p_0\to \infty}}
  2 \frac{u^ 2 t^ 2}{u^ 4+t^ 4} \quad  \xrightarrow[]{t=u} 1  \, .
  \ee
  Therefore, MaxEnt ($\Delta_{RL}=1$) is realized when $t=u$, which corresponds again
  to  the scattering angle $\theta=\pi/2$.
  The indistinguishability of $u$ and $t$ histories is now at the heart of entanglement.
  This also implies that entanglement will not be generated in processes such as
  $e^- \mu^ - \to e^- \mu^ -$ where the same $u/t$ interference
  cannot take place.
  Including electron mass $m_e$ effects, the concurrence $\Delta_{RL}$ reads
\ba
  \label{exactconcurrenceee}
\Bigg|\frac{2tu\left(tu+m_{e}^{2}\frac{(t-u)^{2}}{t+u}\right)}{2m_{e}^{2}(t-u)^{2}(2m_{e}^{2}-2(t+u)+\frac{tu}{t+u})+(t^{4}+u^{4})}\Bigg| \, ,
\ea    
which shows the more powerful result that, for all energies,
the scattering angle $\theta=\pi/2$ (when $t=u$) leads to MaxEnt,
$\Delta_{RL}|_{\theta=\pi/2}=  1$ for all $p_0$.

In the case of incoming particles in an $RR$ product state, no entanglement is generated in the
high-energy limit, since the amplitude is dominated by the final state which also lives
in the $RR$ sector, as required by helicity conservation.
On the other hand, at very low energies the calculation of the concurrence gives
\be
  \label{lowenergyconcurrence}
  \Delta_ {RR}  \xrightarrow[]{|\vec{p}| \ll m_e, \theta = \pi/2}
 1 + O\left(|\vec p|^ 2/m_e^ 2\right) \, .
\ee    
The combination of Eqns.~(\ref{concurrenceee}) and~(\ref{lowenergyconcurrence})
illustrates the remarkable fact that two electrons will get always entangled
at low energies, irrespectively of their initial helicities. It also justifies
that at low energies we easily find entangled fermions and we can
describe their interactions
with effective models such as the Heisenberg model. Electron-electron interaction
is different from all other processes due to the indistinguishability of the particles.

The way in which MaxEnt is generated in QED scattering processes can be studied more thoroughly.
It is indeed possible to show that MaxEnt also arises
in Bhabha scattering and in pair annihilation of electron-positron to two photons. Table \ref{Tab:Ent} shows the MaxEnt states that can be obtained in all tree-level QED processes, both at high and low energies. All processes, with the exception of electron-muon scattering and Compton scattering (photon-electron scattering) can generate maximally entangled states in some energy limit and at a given scattering angle. In two cases, MaxEnt is generated independently of the scattering angle: pair annihilation into photons, and electron-positron annihilation into muons, in both cases at low energy and for an initial state $|RR\rangle$.
It is highly non-trivial that a single coupling, the QED vertex, can take care of
generating entanglement in all these processes, and at the same time
guarantee that 
 if entanglement is present in the initial state, it will be preserved by the interaction.

\begin{table}[t]
\centering
\def\arraystretch{1.5}
\begin{tabular}{c c C{3cm}C{3cm}C{3cm}C{3cm}}
\toprule
 \multicolumn{2}{c}{Process} &\multicolumn{2}{c}{Initial state $|RR\rangle$} & \multicolumn{2}{c}{Initial state $|RL\rangle$} \\
& & High Energy & Low Energy & High Energy & Low Energy \\
 \midrule
Mott scattering & $e^{-}\mu^{-}\rightarrow e^{-}\mu^{-}$ & -- & -- & -- & -- \\
$e^{-} e^{+}$ annihilation & \multirow{2}{*}{$e^{-}e^{+}\rightarrow \mu^{-}\mu^{+}$} & \multirow{2}{*}{--} & \multirow{2}{*}{($\cos\theta|\Phi^{-}\rangle - \sin\theta|\Psi^{+}\rangle)_{\forall \theta} $}& \multirow{2}{*}{$|\Psi^{-}\rangle_{\theta=\pi/2}$} & \multirow{2}{*}{--} \\ [-10pt]
into muons & & & & & \\
M\o ller scattering & $e^{-}e^{-}\rightarrow e^{-}e^{-}$ & -- & $|\Phi^{-}\rangle_{\theta=\pi/2}$ & $|\Psi^{-}\rangle_{\theta=\pi/2}$ & $|\Psi^-\rangle_{\theta=\pi/2}$ \\
Bhabha scattering & $e^{-}e^{+}\rightarrow e^{-}e^{+}$ & -- & -- & $|\Psi^{+}\rangle_{\theta=\pi/2}$ & -- \\
Pair annihilation & $e^{-}e^{+}\rightarrow \gamma\gamma$ & -- & $|\Phi^{-}\rangle_{\forall \theta}$ & $|\Psi^{-}\rangle_{\theta=\pi/2}$ & -- \\
\bottomrule
& &\multicolumn{2}{c}{Initial state $|R+\rangle$} & \multicolumn{2}{c}{Initial state $|R-\rangle$} \\
 & & High Energy & Low Energy & High Energy & Low Energy \\
 \midrule
Compton scattering & $e^{-}\gamma\rightarrow e^{-}\gamma$ & -- & -- & -- & -- \\
\bottomrule
\end{tabular}
\caption{Maximally-entangled states ($\Delta =1$) for tree-level QED processes, both in the high- and low-energy limits. The states are written in terms of the Bell basis, $|\Phi^\pm\rangle\sim|RR\rangle\pm|LL\rangle$ and $|\Psi^{\pm}\rangle\sim|RL\rangle\pm|LR\rangle$. 
For the processes in the upper part of the table, the initial state is expressed in terms of the helicities of the fermions, $R$ and $L$.
For Compton scattering, $e^{-}\gamma\rightarrow e^{-}\gamma$, the initial state is expressed in terms of the helicity of the electron and the polarization of the photon, $|+\rangle$ or $|-\rangle$.
The scattering angle where the entangled state is produced is indicated in the subscript.
A dash indicates that MaxEnt cannot be reached for any value of the scattering angle $\theta$.
}
\label{Tab:Ent}
\end{table}

\section{MaxEnt as a constraining principle}
\label{sec:maxent}

It is tantalizing to turn the discussion upside down and attempt to promote MaxEnt to a
fundamental principle that constraints particle interactions. Following Wheeler's idea of
looking for an Information Theory principle underlying the laws of Nature, we propose to investigate to what extent a MaxEnt principle makes sense in particle physics.
Such principle would guarantee the intrinsically quantum
character of the laws of Nature,
allowing to perform Bell-type experiments 
violating the bounds set by classical physics.
In this formulation, MaxEnt emerges as a purely information-theoretical
principle that can be applied to a variety of problems.

Let us formulate the MaxEnt principle as follows. We shall impose that the 
laws of Nature can generate maximal entanglement in scattering processes of
incoming particles which are not entangled. This should happen in as many
processes as possible. We shall, thus, construct a global figure of merit that
takes into account many processes at a time. To verify the power of such a principle
we shall leave unconstrained the coupling in QED, and analyze the constraint that 
the MaxEnt principle dictates on the coupling. In the next section, we will perform a
similar analysis focusing on the parameters of the weak interaction.

It may be argued that most interactions generate entanglement; however, it is certainly
true that only a limited class of couplings can produce maximal entanglement. 
It is a natural extremization principle which is at play, as it is the case 
in other principles applied to describe Nature. 
Furthermore, maximal entanglement carries the added value that
physics is forced to be non-classical as Bell inequalities are violated. 

We shall further explore this idea in the context
of an unconstrained version of QED (uQED).
This is a theory of fermions and photons, obeying the
 Dirac and Maxwell equations respectively, which can interact via a generic
 vertex that allows violations of rotation and gauge invariance.
For simplicity, we shall still impose the C, P and T discrete symmetries.
While of course this theory is not realized in Nature, our goal is to determine to which extent
imposing MaxEnt constrains this interaction vertex and to verify that QED can be reproduced.

To be more specific, we shall replace the  QED vertex $e\gamma^{\mu}$ with a general
object $eG^\mu$ that can be expanded in a basis of 16 $4\times 4$ matrices.
This unconstrained interaction vertex can be parametrized as
$G^\mu = a_{\mu\nu} \gamma^\nu$, where $a_{\mu\nu}$ are real numbers and $a_{0j}=a_{i0}=0$ for $i,j=1,2,3$ and $ \gamma^\nu$ are the Dirac matrices.
The QED vertex is recovered for
$a_{00}=a_{11}=a_{22}=a_{33}=1$ and $a_{ij}=0$ for $i\not= j$. 
The computation of the amplitude ${\cal M}_{RL\to RL/LR}$
for $e^- e^+ \to \mu^ - \mu^ +$ scattering in uQED at high energies gives
\be
{\cal M}_{RL} \sim (a_{j1}+i a_{j2})(a_{j1} \cos\theta \mp i a_{j2}-a_{j3} \sin\theta ) \, ,
  \ee
  where the $-(+)$ sign corresponds to the $RL\,(LR)$ final state helicities and the sum
  over $j$ is understood.
  By requiring that MaxEnt is realized in the form $|RL\rangle - |LR\rangle$ ($\Delta_{RL}=1$) at $\theta=\pi/2$
  we derive the constraint $(a_{j1}+i a_{j2}) a_{j3}=0$.
  Introducing the positive-defined Hermitian matrix $A_{kl}=a_{kj} a_{lj}$,
  this condition implies $A_{13}=A_{23}=0$, consistent with QED where all $a_ {ij}=0$
  for $i\not=j$.
  While in general it is not justified to assume that MaxEnt in uQED emerges for the same $\theta$
  as in QED, this example shows the constraints which are obtained
  from concurrence maximization.
  
  Let us notice that the uQED formalism allows angular momentum violation. For instance, let us consider the process $e^+ e^- \rightarrow\mu^+ \mu^-$, and take $\theta=\pi/2$ and $a_{13}=a_{23}=a_{33}=0$. If the initial state is $|\Psi^{-}\rangle=\frac{1}{\sqrt{2}}\left(|RL\rangle -|LR\rangle\right)$, i.e. the singlet state, then the final state is proportional to $
 |RL\rangle + |LR\rangle$, and therefore violates angular momentum conservation (see Appendix \ref{app:unconstrainedQED} for more details). 

  The complete application of the MaxEnt principle to uQED requires the computation of all the
  scattering amplitudes in the new theory and then
  the determination of the constraints on
   the $a_{\mu\nu}$ coefficients from the maximization of the concurrences.
  Here we have maximized the sum of the concurrences
  of four different processes: Bhabha and M\o ller scattering,
 $ee\to \gamma\gamma$ and $e^- e^+\to \mu^- \mu^+$,
  accounting for all initial helicity combinations for product states.
  The maximization has been performed both over the  $a_{\mu\nu}$ coefficients
  and over the scattering angle $\theta$.
We find that the solution which maximizes the concurrence is
\be
\left(G^{0},G^{1},G^{2},G^{3}\right)=\left(\pm\gamma^{0},\pm\gamma^{1},\pm\gamma^{2},\pm\gamma^{3}\right) ,
\label{eq:solG}
\ee
where $\gamma^0, \gamma^1, \gamma^2, \gamma^3$ are the gamma matrices in the Dirac representation.
This result shows that QED is indeed
a solution (though not the only one) of requiring MaxEnt for the above subset of scattering processes in uQED.
Some of these solutions are equivalent to QED since a global sign can be absorbed in the electric charge.

The solutions Eq.~(\ref{eq:solG}) are divided into two groups, those related
to QED and those that  are inconsistent
with QED, for instance because they  violate rotation  symmetry.
The latter solutions cannot be ruled out 
since the scattering processes considered here cannot determine
the overall sign of the $\gamma^{\mu}$ matrices, as they always appear in pairs.
Including further
scattering or decay processes which involve three outgoing particles might remove this ambiguity and eliminate the inconsistent solutions.\\
\vspace{-0.3cm}

\section{MaxEnt in the weak interactions}
\label{sec:maxentweak}

The mechanism underlying MaxEnt generation in weak interactions is more subtle, due to the
interplay between vector and axial currents and
between $Z$ and $\gamma$ channels.
The coupling of the $Z$ boson to fermions reads
\ba
i\frac{g}{\cos\theta_{W}}\gamma^{\mu}\left(g^{f}_{V}-g^{f}_{A}\gamma^{5}\right) \, ,
\end{eqnarray}
where the axial and vector couplings
are $g^{f}_{A}=T^{f}_{3}/2$ and $g^{f}_{V}= T^{f}_{3}/2 - Q_{f}\sin^2\theta_{W}$,
and $\theta_W$ is the Weinberg mixing angle.
For electrons and muons, $T_3=-1/2$ and $Q_f=-1$.
Beyond tree level, the Weinberg angle runs with the energy and is scheme dependent.
The PDG average~\cite{Olive:2016xmw} at $Q=m_Z$ in the on-shell scheme is
$\sin^2\theta_W\simeq 0.2234$.
Therefore, the vector coupling $|g_V|$ for electrons is
smaller than the axial one $|g_A|$ by about one order of magnitude.

The effects of the new axial component in the fermion-boson coupling
can be included as follows.
We first consider $e^-e^+ \to \mu^- \mu^+$ scattering  mediated by a $Z$ boson
in the high-energy limit, where $m_Z$ is neglected.
We define the {\it left} and {\it right} couplings as $g_L=g_V+g_A$ and $g_R=g_V-g_A$, which simplifies the
structure of the currents since $j_{\rm in}^{RL}\sim g_R (0,1,i,0)$ and $j_{\rm in}^{LR}\sim g_L (0,1,-i,0)$.
By applying the MaxEnt requirement to the concurrences $\Delta_{RL(LR)}$ (see Appendix~\ref{app:Z} for details) we can then derive
a constraint between the couplings $g_R$ and $g_L$, and the scattering angle $\theta$. 

\begin{figure}[t]
\centering
\includegraphics[width=0.49\textwidth]{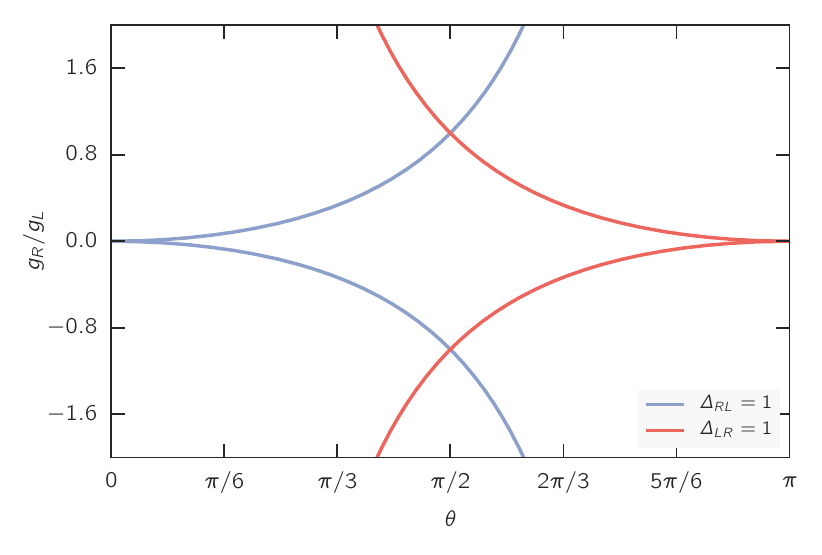}
\includegraphics[width=0.49\columnwidth]{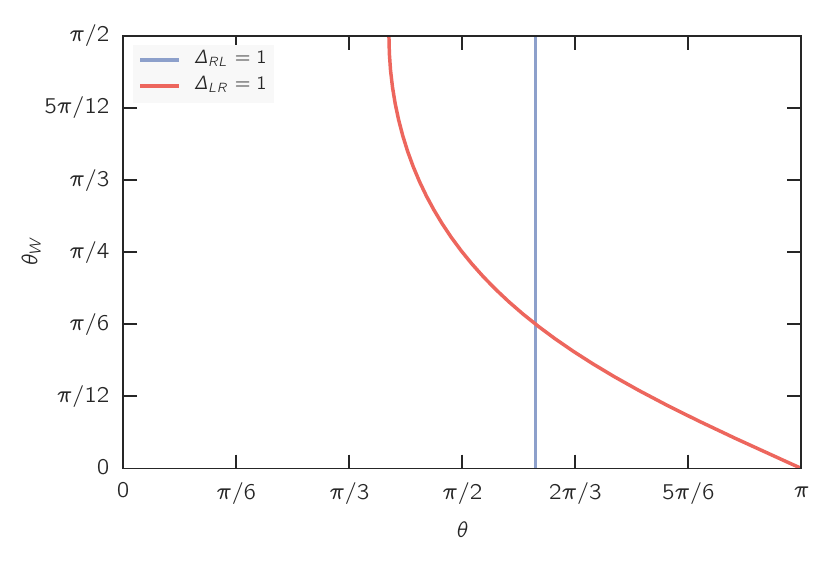}
\vspace{-0.4cm}
\caption{\small Left: Maximal concurrence line as a function of the scattering angle $\theta$ and
  the coupling ratio $g_{R}/g_{L}$ for
  $Z$-mediated $e^{-}e^{+}\to\mu^{-}\mu^{+}$ scattering. Blue line: electron and positron with right- and left-handed initial helicities respectively; red line: electron and positron with left- and right-handed initial helicities. 
 Maximal entanglement is achieved at the same scattering angle $\theta$ for the two initial helicity configurations when the coupling ratio is equal to one, which leads to a Weinberg angle of $\pi/6$.
  Right: Maximum concurrence line for the weak mixing angle $\theta_{W}$ as a function of scattering angle $\theta$ for the process $e^{-}e^{+}\rightarrow\mu^{-}\mu^{+}$, now including also the effects of $Z/\gamma$ interference.
Imposing that MaxEnt is achieved for the same value of
  the scattering angle $\theta$ fixes $\theta_{W}=\pi/6$.
  }
  \label{Fig:emu_Z}
\end{figure}

In the left panel of Fig.~\ref{Fig:emu_Z} we show the maximal concurrence lines ($\Delta=1$) as a function of the scattering
angle $\theta$ and of
  the coupling ratio $g_{R}/g_{L}$ for the two combinations $LR$ and $RL$.
  We find that both concurrences are simultaneously maximized for
  $\theta=\pi/2$, where $g_R=\pm g_L$, that is, either  $g_A=0$
  or $g_V=0$.
  If the axial coupling vanishes $g_A=0$, we  recover the known QED result.
  The $g_V=0$ solution, a vanishing vector coupling, corresponds to
  a Weinberg angle of $\sin^2\theta_W=1/4$, in agreement with the experimental
  value at the $Z$ pole~\cite{Olive:2016xmw} within $\sim 10\%$. 
This result can be traced back to the $Z\to f\bar{f}$ decay,
and indeed the decay of any polarization of the $Z$ particle
gets maximally entangled under the condition $\sin^2\theta_W=1/4$ (see Appendix~\ref{app:Z}). 
Thus scattering processes mediated by a $Z$ inherit the 
 entanglement structure from $Z$ decays.

%

%
   There are two possible explanations for the $\sim 10\%$ discrepancy with respect to the experimental value of the Weinberg angle. On the one hand, this analysis has been performed at first order in perturbation theory; the full MaxEnt analysis should be performed taking into account also higher orders, which modify the amplitudes. On the other hand, it is possible that MaxEnt does not fix this parameter, but only gives us a close value, a first intuition.
   It is however remarkable that requesting MaxEnt simultaneously for the two initial state helicities
  leads either to QED or to a theory which looks surprisingly close to the weak
  interaction.

  Finally, we have studied how the concurrences are modified if we include both the contribution from $\gamma$-exchange and $Z$-exchange in
    $e^{-}e^{+}\to\mu^{-}\mu^{+}$ scattering.
    This is a non-trivial check since the $\gamma$ contribution adds terms to both $RL$ and $LR$, which are independent of $\sin^ 2\theta_W$.    
If we include the effects of the photon/$Z$ interference, the expressions for
the concurrences become more complicated.  In particular, $\Delta_{RL}$ does not depend on the weak mixing angle, but $\Delta_{LR}$ does. Taking the leptonic electric and weak isospin charges $Q=-1$ and $T_{3}=-1/2$, respectively,
we find that (see Appendix~\ref{app:Z})
\begin{eqnarray}
\Delta_{RL}&=&\frac{4\sin^2\theta}{6\cos\theta + 5 (1 +\cos^{2}\theta)}, \\
\Delta_{LR}&=&\frac{\sin^2\theta\sin^2\theta_W}{\cos^4(\theta/2) + 4\sin^4(\theta/2)\sin^4\theta_W}\, .
\end{eqnarray}
Imposing that MaxEnt should be reached for some scattering angles implies that
\begin{eqnarray}
\theta\left(\Delta_{RL}=1\right) &=& \arccos\left(-\frac{1}{3}\right) \ \forall \ \theta_{W}, \\
\theta_{W}\left(\Delta_{LR}=1\right)&=& \arcsin\left(\frac{1}{\sqrt{2}}\cot(\theta/2)\right) \, .
\end{eqnarray}
The two curves are shown in the right panel of Fig.~\ref{Fig:emu_Z}.
If MaxEnt is realized
for the same scattering angle independently of the specific
scattering initial state, then the prediction $\theta_{W}=\pi/6$ readily follows,
consistently with the result that we find by requesting MaxEnt in the decays of
the $Z$ boson into leptons and $e^+e^-\rightarrow\mu^+\mu^-$ scattering mediated by a $Z$ boson.

While the application of MaxEnt to $Z$-boson mediated scattering does not fix completely
the coupling structure of the weak interactions,
as we mentioned its application to $Z$ decay fixes $g_V=0$ and thus
$\sin^2\theta_W=1/4$.
The lack of full predictivity of MaxEnt in the full scattering case
is due to the freedom to choose different angles for MaxEnt depending on the
chirality of the initial particles.

\section{Summary}
\label{sec:summary}

Fundamental interactions generate
entangled states using mechanisms based on indistinguishability, 
consistently with the symmetries of the theory. 
In this work we have explored the relationship between generation of maximally entangled states and high-energy
scattering amplitudes in QED and the weak interactions.
We found that promoting MaxEnt to a fundamental principle in the spirit of
Wheeler's ``it from bit" philosophy
allows one to constrain the coupling structure describing the interactions between fermions
and gauge bosons. As a matter of fact, QED couplings are found to be the solution
to a MaxEnt principle once some global symmetries (C, P and T) are imposed.
We also found that MaxEnt in the weak interactions prefers a weak angle
$\theta_W=\pi/6$, surprisingly close to the SM value.

In this framework, MaxEnt arises as a possible powerful
information principle that can be applied to different processes, bringing in
unexpected constraints on the structure of high-energy
interactions. 
To mention a few possibilities, MaxEnt may provide new insights into the all-order structure of the QED vertex, and may hint at further relations between the parameters of the Standard Model or in new physics beyond it.


\section*{Acknowledgements}
We thank our colleagues L.~\'Alvarez-Gaum\'e\ 
and C.~G\'omez for fruitful discussions.
L.~R. thanks E. Nocera for useful discussions.
A.~C. is grateful to the University of Oxford for hospitality. 


\paragraph{Funding information}
A.~C. and J.~I.~L. acknowledge
support from FIS2015-69167-C2-2-P project.
The work of J.~R. and L.~R. is supported by the ERC Starting Grant ``PDF4BSM''.

\begin{appendix}

  \section{QED scattering amplitudes with helicity dependence}
  \label{app:QED}
  In this appendix we provide explicit results for the calculations of some of the
  relevant  QED scattering amplitudes with explicit helicity dependence.
  Specifically, we consider electron-positron annihilation into muons,
  M{\o}ller scattering, and pair annihilation into photons.
  
\subsection{Electron-positron annihilation into muons: $e^{-}e^{+}\rightarrow\mu^{-}\mu^{+}$}

The first QED process that we consider is electron-positron annihilation into muons,
$e^{-}e^{+}\rightarrow\mu^{-}\mu^{+}$.
At Born level, there is a single Feynman diagram that contributes to
the amplitude, and is shown in Fig.~\ref{Fig:epem_to_mupmum}.
This scattering process is mediated by a virtual photon in the
$s$ channel.
We are interested in computing the
scattering amplitudes for initial and final states
with well defined helicities.
Up to an overall factor which is irrelevant
for the discussion of entanglement, 
for the case of where the initial product state shares the same
helicities, these amplitudes are  given by
\begin{eqnarray}
\mathcal{M}_{|RR\rangle\rightarrow|RR\rangle} &=& -\mathcal{M}_{|RR\rangle\rightarrow|LL\rangle} =-\frac{\lambda}{\mu^{2}+\lambda^{2}}\cos\theta \, ,\nonumber\\
\mathcal{M}_{|RR\rangle\rightarrow|RL\rangle}&=&\mathcal{M}_{|RR\rangle\rightarrow|LR\rangle} = \frac{\lambda}{\sqrt{\mu^{2}+\lambda^{2}}}\sin\theta \, ,\\
\mathcal{M}_{|LL\rangle\rightarrow|RR\rangle} &=& -\mathcal{M}_{|LL\rangle\rightarrow|LL\rangle} =\frac{\lambda}{\mu^{2}+\lambda^{2}}\cos\theta \, ,\nonumber\\
\mathcal{M}_{|LL\rangle\rightarrow|RL\rangle}&=&\mathcal{M}_{|LL\rangle\rightarrow|LR\rangle} = -\frac{\lambda}{\sqrt{\mu^{2}+\lambda^{2}}}\sin\theta \, , \nonumber
\end{eqnarray}
and in the case where the input product state has different helicities,
\begin{eqnarray}
 \label{eq:SIepemmupmm}  
\mathcal{M}_{|RL\rangle\rightarrow|RR\rangle}&=&-\mathcal{M}_{|RL\rangle\rightarrow|LL\rangle}= -\frac{\sin\theta}{\sqrt{\mu^{2}+\lambda^{2}}} \, ,\nonumber\\
\mathcal{M}_{|LR\rangle\rightarrow|RR\rangle}&=&-\mathcal{M}_{|LR\rangle\rightarrow|LL\rangle}= -\frac{\sin\theta}{\sqrt{\mu^{2}+\lambda^{2}}} \, ,\nonumber\\
\mathcal{M}_{|RL\rangle\rightarrow|RL\rangle}&=&-\left(1+\cos\theta\right)\, , \\
\mathcal{M}_{|RL\rangle\rightarrow|LR\rangle}&=& \left(1-\cos\theta\right)\, , \nonumber\\
\mathcal{M}_{|LR\rangle\rightarrow|RL\rangle}&=&\left(1-\cos\theta\right) \, , \nonumber\\
\mathcal{M}_{|LR\rangle\rightarrow|RL\rangle}&=&-\left(1+\cos\theta\right)\, , \nonumber
\end{eqnarray}
where $\theta$ is the scattering angle in the center of mass frame,
$\lambda\equiv m_{e}/m_{\mu}$ is the ratio between the electron and muon mass,
and  $\mu\equiv|\vec{p}|/m_{\mu}$ is the ratio of the momentum of the incoming electrons
over the muon mass.
In the high energy limit, where the center of mass energy of the scattering
is much larger than the muon mass, we have that $\mu\to \infty$.
In this limit it is also of course a very good approximation to assume that
 $\lambda\ll\mu$.
In this limit, we see from Eq.~(\ref{eq:SIepemmupmm}) that all the scattering
amplitudes involving fermions of the same helicity either in the initial or in the final
state are subleading, and the dominant amplitudes are those where both the initial
and final states are composed by particles with different helicity.

Using the concurrence, Eq.~(\ref{concurrence}), to quantify the
amount of entanglement between the helicities of the outgoing particles
present in each of these scattering processes, we obtain that
for a generic value of the center-of-mass energy $\mu$ we have
\begin{eqnarray}
\Delta_{RR}&=& 1\, , \\
\label{eq:DRR}
\Delta_{RL}&=&\frac{\left(\mu^{2}+\lambda^{2}-1\right)\sin^{2}\theta}{\left(\mu^{2}+\lambda^{2}\right)\left(1+\cos^{2}\right)+\sin^{2}\theta}\, .
\label{eq:DRL}
\end{eqnarray}
Therefore, for $RR$ scattering there is always maximally entangled, for any energy.
The same is true for $LL$ scattering.
Note that in the high-energy limit however the contribution to the $RR$ and $LL$ initial
helicity states is suppressed by a factor  $1/\mu$ with respect to the
$RL$ and $LR$ combinations, and thus will contribute much less to the total cross section.

The high-energy limit  for the $RL$ concurrence reads
\begin{eqnarray}
\Delta_{RL}\xrightarrow{\mu\rightarrow\infty} \frac{\sin^{2}\theta}{1+\cos^{2}\theta} +\mathcal{O}\left(\frac{1}{\mu^{2}}\right) \xrightarrow{\theta\rightarrow\pi/2} 1 +\mathcal{O}\left(\frac{1}{\mu^{2}}\right) .
\end{eqnarray}
Therefore, we see that
MaxEnt is realized when the scattering angle is $\theta=\pi/2$,
that is, when the muon-antimuon are scattered perpendicularly to the original
direction of motion of the electron and positron.

\subsection{M{\o}ller scattering: $e^{-}e^{-}\rightarrow e^{-}e^{-}$}

The next process that we consider is M{\o}ller scattering, $e^{-}e^{-}\rightarrow e^{-}e^{-}$,
electron-electron elastic scattering.
The main difference as compared to $e^{+}e^{-}\rightarrow \mu^{+}\mu^{-}$ scattering
is that there is no $s$-channel diagram. Tree-level
M{\o}ller scattering is mediated instead by $t$ and $u$-channel
diagrams, as illustrated in  Fig.~\ref{Fig:moeller}.
This process illustrates a different way of realizing
MaxEnt from an initial product state, by means
of the the interference between $t$ and $u$ channels.

To analyze the way this mechanism for entanglement
generation works, we write the scattering amplitudes
in the helicity basis for the two Feynman diagrams in terms of the $t$ and $u$ Mandelstam variables. The result for initial product states that share
the same helicity reads
\begin{eqnarray}
\mathcal{M}_{|RR\rangle\rightarrow|RR\rangle}&=&\mathcal{M}_{|LL\rangle\rightarrow|LL\rangle}= -\frac{2\left(\left(t+u\right)^{3}-2m_{e}^{2}\left(t^{2}+u^{2}\right)\right)}{tu\left(t+u\right)}\, , 
  \nonumber \\
\mathcal{M}_{|RR\rangle\rightarrow|LL\rangle}&=&\mathcal{M}_{|LL\rangle\rightarrow|RR\rangle}= -\frac{8m_{e}^{2}}{\left(t+u\right)}\, ,
  \nonumber \\
\mathcal{M}_{|RR\rangle\rightarrow|RL\rangle}&=&-\mathcal{M}_{|RR\rangle\rightarrow|LR\rangle}= -\frac{2m_{e}\left(t-u\right)\sqrt{tu\left(4m_{e}^2-(t+u)\right)}}{\left(t+u\right)tu} \, ,
  \\
\mathcal{M}_{|LL\rangle\rightarrow|RL\rangle}&=&-\mathcal{M}_{|LL\rangle\rightarrow|LR\rangle}= -\frac{2m_{e}\left(t-u\right)\sqrt{tu\left(4m_{e}^2-(t+u)\right)}}{\left(t+u\right)tu}\, , \nonumber 
\end{eqnarray}
and for initial product states with opposite helicities we have instead that
\begin{eqnarray}
\mathcal{M}_{|RL\rangle\rightarrow|RR\rangle}&=&\mathcal{M}_{|RL\rangle\rightarrow|LL\rangle} = \frac{2m_{e}\left(t-u\right)\sqrt{tu\left(4m_{e}^2-(t+u)\right)}}{\left(t+u\right)tu}\,,
  \nonumber \\
\mathcal{M}_{|LR\rangle\rightarrow|RR\rangle}&=&\mathcal{M}_{|LR\rangle\rightarrow|LL\rangle} = -\frac{2m_{e}\left(t-u\right)\sqrt{tu\left(4m_{e}^2-(t+u)\right)}}{\left(t+u\right)tu}\,,
  \nonumber \\
\mathcal{M}_{|RL\rangle\rightarrow|RL\rangle}&=&\mathcal{M}_{|LR\rangle\rightarrow|LR\rangle}= -\frac{2\left(2m_{e}^{2}\left(t-u\right)+u\left(t+u\right)\right)}{t\left(t+u\right)}\,, 
   \\
\mathcal{M}_{|RL\rangle\rightarrow|LR\rangle}&=& \mathcal{M}_{|LR\rangle\rightarrow|RL\rangle} =-\frac{2\left(2m_{e}^{2}\left(t-u\right)-t\left(t+u\right)\right)}{u\left(t+u\right)}\, . \nonumber
\end{eqnarray}
Using these helicity scattering amplitudes, we can now compute the concurrences for the two
relevant cases: an incoming product state where the two fermions share the same helicity,
or where the incoming fermions have opposite helicities.
The values of these concurrences, valid for any value of the center-of-mass energy
(not only in the high-energy approximation) turn out to be given by:
\begin{eqnarray}
\Delta_{RR}&=&\Bigg|\frac{2m_{e}^{2}tu\left(\left(3t+u\right)\left(t+3u\right)-4m_{e}^{2}\left(t+u\right)\right)}{\left(t+u\right)^{5}-2m_{e}^{2}\left(2t^{4}+5t^{3}u+2t^{2}u^{2}+5tu^{3}+2u^{4}\right)+ 4m_{e}^{4}\left(t+u\right)\left(t^{2}+u^{2}\right)}\Bigg|\, , \\
\Delta_{RL}&=&\Bigg|\frac{2tu\left(tu\left(t+u\right)+m_{e}^{2}\left(t-u\right)^{2}\right)}{2m_{e}^{2}\left(t-u\right)^{2}\left(2m_{e}^{2}\left(t+u\right)-\left(2t^{2}+3tu+2u^{2}\right)\right)+ \left(t+u\right)\left(t^{4}+u^{4}\right)}\Bigg| \, .
\end{eqnarray}
Recalling the fact that the kinematical condition $u=t$ corresponds to
an scattering angle of $\theta=\pi/2$, we obtain that for this configuration
\begin{eqnarray}
\Delta_{RR}&\xrightarrow{u=t}& \frac{m_e^2\left(m_{e}^2 - 2  t\right)}{ m_{e}^4 - 2 m_{e}^2 t + 2 t^2} \xrightarrow{t\ll m_{e}} 1 +\mathcal{O}\left(\frac{t^{2}}{m_{e}^{4}}\right) \, ,\\
\Delta_{RL}&\xrightarrow{u=t}& 1 \, .
\end{eqnarray}
Therefore, we find that in M{\o}ller scattering MaxEnt is realized at a
scattering angle $\theta=\pi/2$ for all energies provided that the incoming
particles have opposite helicities.
On the other hand, the same analysis also implies
that for product states composed by same-helicity
particles, MaxEnt is only realized at very low energies, well below the electron mass.

\subsection{Pair annihilation to photons: $e^-e^+\rightarrow\gamma\gamma$}

\begin{figure}[t]
\centering
    {	\includegraphics[width=.25\columnwidth]{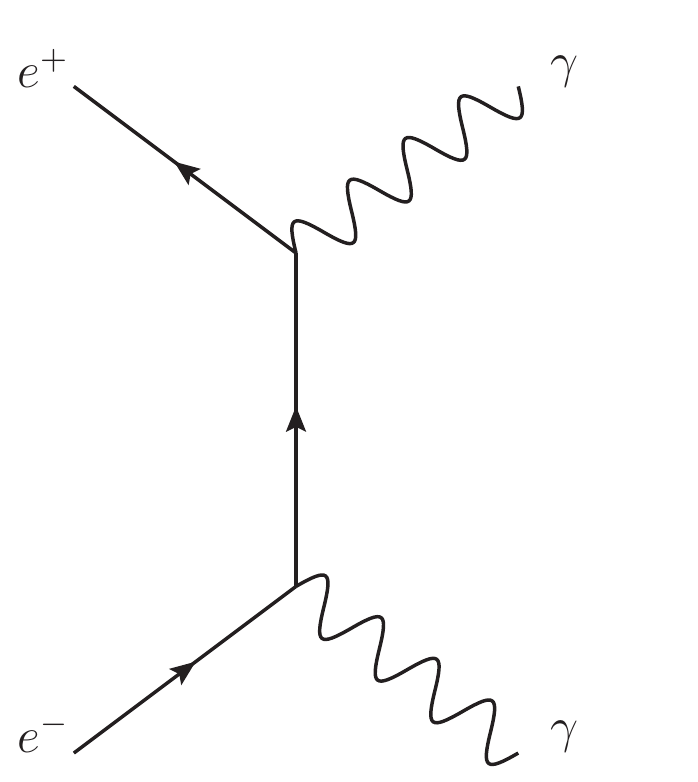}}
  {\includegraphics[width=.25\columnwidth]{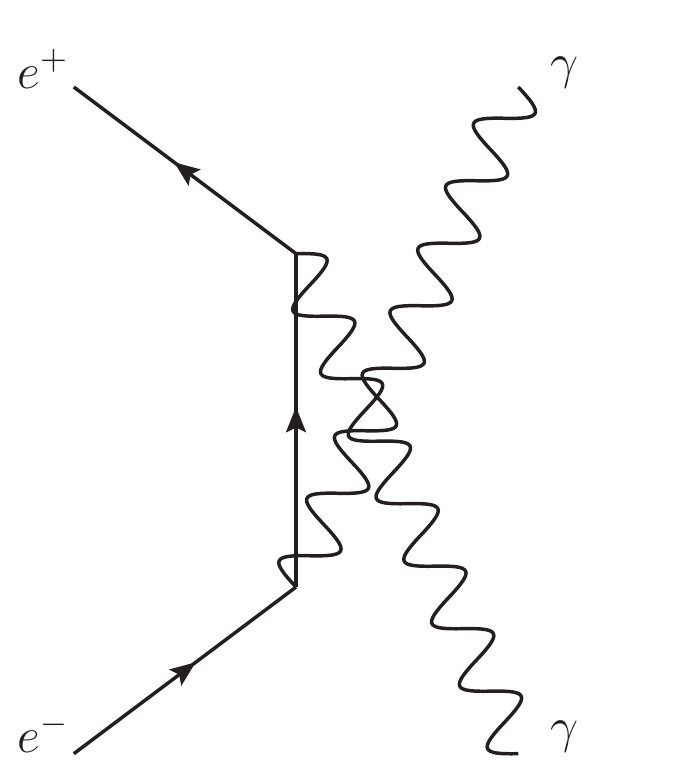}}
\vspace{-0.2cm}
\caption{\small  Feynman diagrams for
  pair annihilation to photons, $e^-e^-\to \gamma \gamma$,
  in the   $t$ (left) and $u$ (right) channels.
 }
  \label{Fig:pair}
\end{figure}

This process is another example of indistinguishability as a basic source of entanglement.
Note that pair annihilation to two photons is also described
by the combination of  $t$ and $u$ channels (see Fig.~\ref{Fig:pair}), as was
the case for M{\o}ller scattering.
The final-state photons can be
expressed in the circular polarization basis, defined as follows:
\begin{eqnarray}
&\epsilon_{\lambda}(\theta,\phi)=\frac{\lambda}{\sqrt{2}}\left(0,\cos\theta\cos\phi + i\lambda\sin\phi,\cos\theta\sin\phi -i\lambda\cos\phi, -\sin\theta\right),& \\
&|R\rangle\equiv \epsilon_{\lambda=+1}(\theta,\phi) \qquad |L\rangle\equiv \epsilon_{\lambda=-1}(\theta,\phi)&
\end{eqnarray}
The helicity scattering
amplitudes in terms of the Mandelstam variables $u$ and $t$ are:
\begin{eqnarray}
\mathcal{M}_{|RR\rangle\rightarrow|RR\rangle}&=&-\mathcal{M}_{|LL\rangle\rightarrow|LL\rangle}= -\frac{m(t+u-2m^2)}{(m^2-t)(m^2-u)}\left(\sqrt{2m^2-(t+u)}-\sqrt{-2m^2-(t+u)}\right)\,, \nonumber\\
\mathcal{M}_{|RR\rangle\rightarrow|LL\rangle}&=&-\mathcal{M}_{|LL\rangle\rightarrow|RR\rangle}= \frac{m(t+u-2m^2)}{(m^2-t)(m^2-u)}\left(\sqrt{2m^2-(t+u)}+\sqrt{-2m^2-(t+u)}\right)\,, \nonumber\\
\mathcal{M}_{|RR\rangle\rightarrow|RL\rangle}&=&\mathcal{M}_{|RR\rangle\rightarrow|LR\rangle}= -\frac{4m\left(m^4-tu\right)}{(m^2-t)(m^2-u)\sqrt{-2m^2-(t+u)}}\,, \\
\mathcal{M}_{|LL\rangle\rightarrow|RL\rangle}&=&\mathcal{M}_{|LL\rangle\rightarrow|LR\rangle}=
-\mathcal{M}_{|RR\rangle\rightarrow|RL\rangle}\,, \nonumber
\end{eqnarray}
and
\begin{eqnarray}
\mathcal{M}_{|RL\rangle\rightarrow|RL\rangle}&=&\mathcal{M}_{|RL\rangle\rightarrow|LR\rangle}=\mathcal{M}_{|LR\rangle\rightarrow|RL\rangle}=\mathcal{M}_{|LR\rangle\rightarrow|LR\rangle}=0\, , \\
\mathcal{M}_{|RL\rangle\rightarrow|RL\rangle}&=&\mathcal{M}_{|LR\rangle\rightarrow|LR\rangle}=\sqrt{\frac{tu-m^4}{(t+u)^2-4m^4}}\frac{\sqrt{(t+u)^2-4m^4}(u-t)-(4m^4-(t+u)^2)}{(m^2-t)(m^2-u)\sqrt{1+\frac{4m^2}{(t+u)-2m^2}}}\, , \nonumber \\
\mathcal{M}_{|RL\rangle\rightarrow|LR\rangle}&=&\mathcal{M}_{|LR\rangle\rightarrow|RL\rangle}=\sqrt{\frac{tu-m^4}{(t+u)^2-4m^4}}\frac{\sqrt{(t+u)^2-4m^4}(u-t)+(4m^4-(t+u)^2)}{(m^2-t)(m^2-u)\sqrt{1+\frac{4m^2}{(t+u)-2m^2}}}\, . \nonumber
\end{eqnarray}
The corresponding concurrences are then given by
\begin{eqnarray}
\Delta_{RR}&=&\Bigg|\frac{2m^2\left(4m^6-4m^4(t+u)-2m^2(t-u)^2+(t+u)^3\right)-8t^2u^2}{4m^4\left(2m^4+2m^2(t+u)-(t^2+6tu+u^2)\right)+t^4+4t^3u+14t^2u^2+4tu^3+u^4}\Bigg|, \nonumber \\
\Delta_{RL}&=&\frac{2(tu-m^4)}{t^2+u^2-2m^4}.
\end{eqnarray}

As in the case of M{\o}ller scattering, if $\theta=\pi/2$ (corresponding to $t=u$) MaxEnt is realized for all energies when the initial particles have opposite helicities, while MaxEnt arises only for small momentum
transfers $t$ in the case in which the initial-state particles share the same helicity,
\begin{eqnarray}
\Delta_{RR}&\xrightarrow{u=t}&\frac{(m^2-t)}{m^2+3t} \xrightarrow{t\ll m_{e}} 1 + \mathcal{O}\left(\frac{t}{m_{e}^{2}}\right), \\
\Delta_{RL}&\xrightarrow{u=t}& 1 \ \forall\, t.
\end{eqnarray}

\section{Unconstrained QED}
\label{app:unconstrainedQED}

In this appendix we provide more information about unconstrained QED (uQED) and about the
form of the helicity scattering amplitudes
computed in this hypothetical theory.
The Dirac matrices $\gamma^{\mu}$ form the Clifford algebra of the $4\times 4$ matrices; $\{\gamma^{\mu},\gamma^{\nu}\}=2g^{\mu\nu}$ if the metric is $(+,-,-,-)$.
The complexification of the Clifford algebra $C\ell_{1,3}(\mathbf{R})$,  $C\ell_{1,3}(\mathbf{R})_{\mathbf{C}}$, is isomorphic to the algebra of $4\times 4$ complex matrices.
Therefore, it is possible to express any
general $4\times 4$ complex matrix as
\begin{eqnarray}
G^{\mu}=c_{1}^{\mu}\mathbb{I}+c_{2}^{\mu\nu}\gamma_{\nu}+ic_{3}^{\mu}\gamma^{5}+ c_{4}^{\mu\nu}\gamma^{5}\gamma_{\nu}+c_{5}^{\mu\nu\rho}\sigma_{\nu\rho},
\label{eq:generalG}
\end{eqnarray}
where the expansion coefficients are real-valued, $c_{i}\in\mathbb{R}$, and $\gamma^{5}=i\gamma^{0}\gamma^{1}\gamma^{2}\gamma^{3}$, $\sigma^{\nu\rho}=-\frac{i}{2}\left[\gamma^{\nu},\gamma^{\rho}\right]$.
The hypothetical theory of unconstrained QED is constructed by
replacing the QED vertex $-ie\gamma^{\mu}$ by the general $4\times 4$
complex matrices $-ieG^{\mu}$.
To simplify the analysis, we first impose the conservation of
the C, P and T discrete symmetries,
which leads to $c_{1}^{\mu}=c_{3}^{\mu}=c_{4}^{\mu\nu}=c_{5}^{\mu\nu\rho}=0$ and $c_{2}^{\mu\nu}\equiv a^{\mu\nu}$ with $a^{i0}=a^{0j}=0$ for $i,j=1,2,3$.\\
With these assumptions for the electron-photon interaction vertex, when computing the amplitudes at high energy limit for the process $e^{+}e^{-}\rightarrow\mu^{+}\mu^{-}$ using and restricting
the particles to be in the $XZ$ plane one obtains the following results for incoming $|RL\rangle$:
\begin{eqnarray}
\mathcal{M}_{|RL\rangle\rightarrow|RL\rangle}&=& -a_{j2}^2 - a_{j1}^2\cos\theta + a_{j1} a_{j3}\sin\theta + i\left(a_{j1} a_{j2} (1-\cos\theta) + a_{j2} a_{j3}\sin\theta\right)\, ,\nonumber \\
\mathcal{M}_{|RL\rangle\rightarrow|LR\rangle}&=& a_{j2}^2 - a_{j1}^2\cos\theta + a_{j1} a_{j3}\sin\theta - i\left(a_{j1} a_{j2} (1+\cos\theta) - a_{j2} a_{j3}\sin\theta\right)\, , 
\label{eq:uemu}
\end{eqnarray}
while all other scattering amplitudes vanish.

The two possible final states that maximize the concurrence, that is, that realize MaxEnt
are given by $|RL\rangle\pm|LR\rangle$,
and therefore
$\mathcal{M}_{|RL\rangle\rightarrow|RL\rangle}=\pm\mathcal{M}_{|RL\rangle\rightarrow|LR\rangle}$.
Requiring a final state that satisfies the maximal entanglement principle
we find that for a scattering angle of $\theta=\pi/2$ the following conditions must be satisfied
\begin{eqnarray}
\begin{array}{rcl}
a_{j2}^2 - i a_{j1}a_{j2} = 0 & \longrightarrow A_{22} = A_{12} = 0 & \mathrm{if} \ \mathcal{M}_{|RL\rangle}=\mathcal{M}_{|LR\rangle} \quad \mathrm{or}\\
a_{j1}a_{j3} + i a_{j2}a_{j3} = 0& \longrightarrow A_{13}=A_{23} = 0 & \mathrm{if} \ \mathcal{M}_{|RL\rangle}=-\mathcal{M}_{|LR\rangle},
\end{array}
\end{eqnarray}
where $A_{kl}\equiv a_{jk}a_{jl}$ is a positive definite matrix.
It is also possible to redo
the same analysis but now requiring the scattered
particles to lie in the $YZ$ and $XY$ planes
respectively.
If the motion of the initial particles
takes place in the $Y$ axis and the outgoing scattered particles
lie in the $YZ$ plane, the corresponding scattering
amplitudes become:
\begin{eqnarray}
\mathcal{M}_{|RL\rangle\rightarrow|RL\rangle}&=& -a_{j1}^2 - a_{j2}^2\cos\theta + a_{j2} a_{j3}\sin\theta - i\left(a_{j1} a_{i2} (1-\cos\theta) + a_{j1} a_{j3}\sin\theta\right),\nonumber \\
\mathcal{M}_{|RL\rangle\rightarrow|LR\rangle}&=& a_{j1}^2 - a_{j2}^2\cos\theta + a_{j2} a_{j3}\sin\theta + i\left(a_{j1} a_{j2} (1+\cos\theta) - a_{j1} a_{j3}\sin\theta\right). 
\end{eqnarray}
If now we request MaxEnt to be realized at an scattering
angle of $\theta=\pi/2$, one finds that
\begin{eqnarray}
\begin{array}{rcl}
a_{j1}^2 + i a_{j1}a_{j2} = 0 & \longrightarrow A_{11} = A_{12} = 0 & \mathrm{if} \ \mathcal{M}_{|RL\rangle}=\mathcal{M}_{|LR\rangle} \quad \mathrm{or}\\
a_{j2}a_{j3} - i a_{j1}a_{j3} = 0& \longrightarrow A_{23}=A_{13} = 0 & \mathrm{if} \ \mathcal{M}_{|RL\rangle}=-\mathcal{M}_{|LR\rangle}.
\end{array}
\end{eqnarray}
For incoming particles in $X$ axis and outgoing in the $XY$ plane, the results read instead
\begin{eqnarray}
\mathcal{M}_{|RL\rangle\rightarrow|RL\rangle}&=& -a_{j3}^2 - a_{j2}^2\cos\phi + a_{j1} a_{j2}\sin\phi + i\left(a_{j2} a_{i3} (1-\cos\phi) + a_{j1} a_{j3}\sin\phi\right),\nonumber \\
\mathcal{M}_{|RL\rangle\rightarrow|LR\rangle}&=& -a_{j3}^2 + a_{j2}^2\cos\phi - a_{j1} a_{j2}\sin\phi + i\left(a_{j2} a_{j3} (1+\cos\phi) - a_{j1} a_{j3}\sin\phi\right). 
\end{eqnarray}
where $\phi$ is the azimuthal angle that goes from 0 to $2\pi$. Fixing $\phi=\pi/2$ we get another set
of conditions
\begin{eqnarray}
\begin{array}{rcl}
a_{j1}a_{j2} + i a_{j1}a_{j3} = 0 & \longrightarrow A_{12} = A_{13} = 0 & \mathrm{if} \ \mathcal{M}_{|RL\rangle}=\mathcal{M}_{|LR\rangle} \quad \mathrm{or}\\
a_{j3}^2 - i a_{j2}a_{j3} = 0& \longrightarrow A_{23}=A_{33} = 0 & \mathrm{if} \,
\mathcal{M}_{|RL\rangle}=-\mathcal{M}_{|LR\rangle} \, .
\end{array}
\end{eqnarray}

A crucial property of  entanglement is that it should be
invariant under local unitary transformations like rotations.
For this reason,
it is possible to obtain the $|RL\rangle + |LR\rangle$ state in one plane and $|RL\rangle -|LR\rangle$ state
in another, but both for the same scattering angle because of isometry.
Therefore, there are a finite number of possible solutions that satisfy the above constraints:
the one which corresponds to QED is the $|RL\rangle - |LR\rangle$ state for $XZ$ and $YZ$ plane and $|RL\rangle + |LR\rangle$ state for $XY$ plane.
While in this example we have imposed that MaxEnt is realized for
specific choices of the scattering angles $\theta,\phi=\pi/2$, it is conceivable that additional
constraints could be obtained by exploiting the information contained in other scattering angles.

From this specific example among the list of processes that
we have analyzed in unconstrained QED, one can also observe that it is not possible
to distinguish the overall sign of the  $a_{ij}$ coefficients,
as they always appear squared or multiplied in pairs.
Other processes, involving for example a final state with three particles, might be necessary
in order to resolve this degeneracy.

Notice also that uQED allows angular momentum violation. Let us take as example the amplitudes for the process $e^+ e^- \rightarrow\mu^+ \mu^-$ in the $XZ$ plane. The results for an initial state $|RL\rangle$ are collected in Eq.~\eqref{eq:uemu}, whereas the corresponding amplitudes for an initial state $|LR\rangle$ read
\begin{eqnarray}
\mathcal{M}_{|LR\rangle\rightarrow|RL\rangle}&=& -a_{j2}^2 + a_{j1}^2\cos\theta - a_{j1} a_{j3}\sin\theta + i\left(-a_{j1} a_{j2} (1+\cos\theta) + a_{j2} a_{j3}\sin\theta\right)\, ,\nonumber \\
\mathcal{M}_{|LR\rangle\rightarrow|LR\rangle}&=& a_{j2}^2 + a_{j1}^2\cos\theta - a_{j1} a_{j3}\sin\theta + i\left(a_{j1} a_{j2} (-1+\cos\theta) - a_{j2} a_{j3}\sin\theta\right)\, . 
\end{eqnarray}
Therefore, if the initial state is $|\Psi^{-}\rangle=\frac{1}{\sqrt{2}}\left(|RL\rangle -|LR\rangle\right)$, i.e. the singlet state, then the final state $|\psi\rangle_{\rm{final}}$ becomes
  \begin{eqnarray}
  |\psi\rangle_{\rm{final}}&\sim& \mathcal{M}_{|RL\rangle\rightarrow|RL\rangle}|RL\rangle + \mathcal{M}_{|RL\rangle\rightarrow|LR\rangle}|LR\rangle - \left(\mathcal{M}_{|LR\rangle\rightarrow|RL\rangle}|RL\rangle + \mathcal{M}_{|LR\rangle\rightarrow|LR\rangle}|LR\rangle\right) \nonumber\\
  &\sim& \left(-\sum_{j}a_{j1}^2\cos\theta + \sum_{j}a_{j1}a_{j3}\sin\theta\right)\left(|RL\rangle-|LR\rangle\right) + i\sum_{j}a_{j1}a_{j2}\left(|RL\rangle + |LR\rangle\right),
  \end{eqnarray}
  which, in general, is not a singlet state: as long as $\sum_{j}a_{j1}a_{j2}\neq 0$, angular momentum is violated in this process $\forall\theta$.

\section{Electroweak processes with helicity dependence}
\label{app:Z}

Finally, we provide in this appendix explicit expressions for tree-level
electroweak processes with helicity dependence, first
for the $Z$ decay into $e^{-}e^{+}$, and then for
the $e^{-}e^{+}\rightarrow\mu^{-}\mu^{+}$ process, mediated by a $Z$ boson and finally including the effects of $Z/\gamma$ interference.

\subsection{$Z$ decay into $e^{-}e^{+}$}

We now analyze the helicity structure of $Z$ boson decay to $e^{-}e^{+}$. As $Z$ is a massive particle, it has three possible polarizations: right- and left-handed circular polarizations,
and longitudinal polarization, which we will denote as $|0\rangle$.
As $m_{e}\ll m_Z$ we can neglect the electron mass.
The non-vanishing  helicity amplitudes for this decay process are:
\begin{eqnarray}
\mathcal{M}_{|0\rangle\rightarrow|RL\rangle}&=& g_{R}m_Z\sin\theta \, , \nonumber\\
\mathcal{M}_{|0\rangle\rightarrow|LR\rangle}&=& g_{L}m_Z\sin\theta \, , \nonumber\\
\mathcal{M}_{|R\rangle\rightarrow|RL\rangle}&=& g_{R}m_Z\sqrt{2}\sin^{2}(\theta/2) \, , \nonumber\\
\mathcal{M}_{|R\rangle\rightarrow|LR\rangle}&=& -g_{L}m_Z\sqrt{2}\cos^{2}(\theta/2)\, , \\
\mathcal{M}_{|L\rangle\rightarrow|RL\rangle}&=& g_{R}m_Z\sqrt{2}\cos^{2}(\theta/2) \, ,\nonumber\\
\mathcal{M}_{|L\rangle\rightarrow|LR\rangle}&=& -g_{L}m_Z\sqrt{2}\sin^{2}  (\theta/2) \nonumber \, ,
\end{eqnarray}
where we have defined $g_{R}=(g_{V}-g_{A})/2$ and $g_{L}=(g_{V}+g_{A})/2$. 

If the $Z$ boson is longitudinally polarized, the concurrence of the final leptons becomes
\begin{eqnarray}
\Delta_{0}&=&\frac{2|g_{L}g_{R}|}{g_{L}^{2}+g_{R}^{2}} \, ,
\end{eqnarray}
Then one can see that the leptons pair is maximally entangled
provided that $|g_{L}|=|g_{R}|$, {\it i.e.} if $g_{A}=0$ or $g_{V}=0$.
As $g_{A}=T_{3}/2\neq 0$ the only possible solution is $g_{V}=0$ which leads to
the value $\theta_{W}=\pi/6$ of the weak mixing angle.

For a $Z$ boson initially polarized with either a right- or left-handed
polarization, the concurrence becomes instead
\begin{eqnarray}
\Delta_{R}&=&\frac{2|g_{L}g_{R}|\sin^{2}\theta}{|2\left(g_{L}^{2}-g_{R}^{2}\right)\cos\theta +\left(g_{L}^{2}+g_{R}^{2}\right)(1+\cos^{2}\theta)|}\, , \\
\Delta_{L}&=& \frac{2|g_{L}g_{R}|\sin^{2}\theta}{|2\left(g_{L}^{2}-g_{R}^{2}\right)\cos\theta -\left(g_{L}^{2}+g_{R}^{2}\right)(1+\cos^{2}\theta)|}\, .
\end{eqnarray}
We already showed in Fig.~\ref{Fig:emu_Z} the dependence of $g_{R}/g_{L}$ for the maximum concurrence as a function of scattering angle $\theta$. As long as $g_{R}/g_{L}=\pm\cot^{2}(\theta/2)$, for an initial right-handed polarization, or $g_{R}/g_{L}=\pm\tan^{2}(\theta/2)$, for an initial left-handed polarization, MaxEnt is achieved. However, if we assume the same relation between $g_{R}/g_{L}$ independently of the initial polarization, then only one solution is possible: $g_{R}/g_{L}=\pm 1$, i.e. the same solution as for longitudinal polarization, $g_V=0$ or equivalently $\theta_{W}=\pi/6$.


\subsection{$e^-e^+ \to \mu^- \mu^+$ mediated by $Z$ boson}

Let us consider $e^-e^+ \to \mu^- \mu^+$ scattering  mediated by a $Z$ boson in the high energy limit, where $m_Z$ is neglected.
The resulting scattering amplitudes are:
\ba
  \begin{array}{l}
 {\cal M}_{LR} \sim(1+\cos\theta) g_{L}^2\ket{LR} + (-1+\cos\theta) g_L g_R \ket{RL} \, , \\[0.2cm]
    {\cal M}_{RL} \sim(-1+\cos\theta) g_R g_L \ket{LR} 
  + (1+\cos\theta) g_R^ 2\ket{RL} \, ,
  \end{array}
\ea
where $g_L=g_V+g_A$ and $g_R=g_V-g_A$, 
and their concurrences for $|\vec p| \gg m_Z$ read:
\be
  \label{concurrenceZ}
   \Delta_{LR\,(RL)}\simeq  
   \frac{ \sin^2\theta |g_L g_R|}{2(c^4 g_L^2 + s^4 g_R^2)}\,\,
   \left( \frac{ \sin^2\theta |g_L g_R|}{2(s^4 g_L^2 + c^4 g_R^2)} \right) \,   ,
\ee
where $c=\cos \theta/2$ and $s=\sin \theta/2$ depend on the scattering angle $\theta$.
Applying the MaxEnt requirement to the above concurrences we find
$c^2 g_L \pm s^2 g_R=0$ ($s^2 g_L \pm c^2 g_R=0$) for the $LR$~($RL$)
initial states.
Note that in general concurrence
maximization occurs for different values of $\theta$ for each initial state.

Both concurrences are simultaneously maximized for
  $\theta=\pi/2$, where $g_R=\pm g_L$. Therefore, either the axial coupling is zero, recovering the known QED result, or the vector coupling is zero, leading to a Weinberg angle of $\sin^2\theta_W=1/4$. 

\subsection{$e^{-}e^{+}\rightarrow\mu^{-}\mu^{+}$ with $Z/\gamma$ interference}

We now revisit the $e^{-}e^{+}\rightarrow\mu^{-}\mu^{+}$ scattering processes, now
taking into account the effects of $Z/\gamma$ interference.
Given that $m_{e},m_{\mu}\ll m_Z$, we can neglect the masses of both leptons.
In this process, the amplitudes with equal initial helicities vanish, while
the scattering amplitudes for opposite initial helicity configurations are given by:
\begin{eqnarray}
\mathcal{M}_{|RL\rangle\rightarrow|RL\rangle}&=& -\left(\frac{4\mu^{2} g_{R}^{2} }{\left(4\mu^{2}-1\right)}\sec^{2}\theta_{W} +Q^{2}\sin^{2}\theta_{W} \right)\left(1+\cos\theta\right)\, , \nonumber\\
\mathcal{M}_{|RL\rangle\rightarrow|LR\rangle}&=& \left(\frac{4\mu^{2} g_{R} g_{L} }{\left(4\mu^{2}-1\right)}\sec^{2}\theta_{W}+Q^{2}\sin^{2}\theta_{W}\right)\left(1-\cos\theta\right)\, ,  
\label{eq:Memuweak}\\
\mathcal{M}_{|LR\rangle\rightarrow|RL\rangle}&=& \mathcal{M}_{|RL\rangle\rightarrow|\substack{LR}\rangle} \left(g_{R}\leftrightarrow g_{L}\right)  \, ,\nonumber\\
\mathcal{M}_{|LR\rangle\rightarrow|LR\rangle}&=& \mathcal{M}_{|RL\rangle\rightarrow|\substack{RL}\rangle} \left(g_{R}\leftrightarrow g_{L}\right) \,, \nonumber
\end{eqnarray}
where we have defined $\mu\equiv|\vec{p}|/m_Z$.

The purely weak scattering
process $e^{-}e^{+}\rightarrow\mu^{-}\mu^{+}$, {\it i.e.},
where the two currents exchange a $Z$ boson instead of a photon like in QED, can be obtained if we set $Q=0$ in the amplitudes of Eq.~\eqref{eq:Memuweak}.

The introduction of the photon channel complicates the expressions for the concurrences.
They simplify if we express them in terms of $Q$ and $T_{3}$, in which case we find
\begin{eqnarray}
\Delta_{RL}&=&\frac{2Q\left(Q-T_{3}\right)\sin^{2}\theta}{2\left(2Q-T_{3}\right)T_{3}\cos\theta +\left(\left(Q-T_{3}\right)^{2}+Q^{2}\right)\left(1+\cos^{2}\theta\right)},\\
\Delta_{LR}&=&\frac{Q\left(Q-T_{3}\right)\sin^{2}\theta\sin^{2}\theta_{W} \left(T_{3}^{2}+Q^{2}\sin^{2}\theta_{W}-2QT_{3}\sin^{2}\theta_{W}\right)}{2Q^{2}\left(Q-T_{3}\right)^{2}\sin^{4}(\theta/2)\sin^{4}\theta_{W}+ 2\left(T_{3}^{2}+Q^{2}\sin^{2}\theta_{W}-2QT_{3}\sin^{2}\theta_{W}\right)^{2}\cos^{4}(\theta/2)}. \nonumber
\end{eqnarray}
Note that $\Delta_{RL}$ does not depend on the weak mixing angle, but $\Delta_{LR}$ does.


\end{appendix}


\end{document}